\documentclass[pre,aps,floatfix,twocolumn,tightenlines,superscriptaddress,amsmath,amssymb,showkeys,nofootinbib,10pt]{revtex4-2}
\usepackage[english]{babel}
\usepackage[utf8]{inputenc}
\usepackage{xcolor}
\usepackage{graphicx}
\usepackage{tabularx}
\usepackage{hyperref}
\hypersetup{colorlinks=true, linktoc=all, linkcolor=blue, linktocpage, citecolor=blue}
\DeclareMathOperator\erfi{Erfi}
\DeclareMathOperator\erf{Erf}
\DeclareMathOperator\erfc{Erfc}

\begin{document}
\title{Coarsening in the Persistent Voter Model: analytical results}

\author{R. G. de Almeida}
\email{rafael_{}almeida@id.uff.br}
\affiliation{Instituto de Física, Universidade Federal Fluminense, Niterói, Brazil} 

\author{J. J. Arenzon}
\email{arenzon@if.ufrgs.br}
\affiliation{Instituto de Física, Universidade Federal do Rio Grande do Sul, CEP 91501-970, Porto Alegre - RS, Brazil} 
\affiliation{Instituto Nacional de Ciência e Tecnologia - Sistemas Complexos, Rio de Janeiro RJ, Brazil}

\author{F. Corberi}
\email{fcorberi@unisa.it}
\affiliation{Dipartimento di Fisica, Universit\`a di Salerno, Via Giovanni Paolo II 132, 84084 Fisciano (SA), Italy}
\affiliation{INFN Sezione di Napoli, Gruppo collegato di Salerno, Italy}

\author{W. G. Dantas}
\email{wgdantas@id.uff.br}
\affiliation{Departamento de Ciências Exatas, EEIMVR, Universidade Federal Fluminense, CEP 27255-125, Volta Redonda - RJ, Brazil}

\author{L. Smaldone}
\email{lsmaldone@unisa.it}
\affiliation{Dipartimento di Fisica, Universit\`a di Salerno, Via Giovanni Paolo II 132, 84084 Fisciano (SA), Italy}
\affiliation{INFN Sezione di Napoli, Gruppo collegato di Salerno, Italy}

\begin{abstract}
We investigate the coarsening dynamics of a simplified version of the persistent voter model in which an agent can become a zealot -- i.e. resistent to change opinion --  at each step, based on interactions with its nearest neighbors. We show that such a model captures the main features of the original, non-Markovian, persistent voter model. We derive the governing equations for the one-point and two-point correlation functions. As these equations do not form a closed set, we employ approximate closure schemes, whose validity was confirmed through numerical simulations. Analytical solutions to these equations are obtained and well agree with the numerical results. 
\end{abstract}

\maketitle

\section{Introduction}

Coarsening is a hallmark of systems undergoing an ordering process~\cite{Bray94} and can occur in liquid crystals~\cite{SiArDiBrCuMaAlPi08,Almeida23}, superconductors~\cite{PrFiHoCa08}, many biological processes~\cite{Mcnally17,Smerlak18,Siteur22,Grober23}, 
among several other examples~\cite{Scheucher88,Cox86}. 
These systems are partitioned into domains whose average area increases in time.
In the scaling regime of the Model A universality class, i.e., scalar, non-conserved order parameter models, there is a single length scale (e.g., the average radius of the domains), that increases as $\ell(t)\sim t^{1/2}$.
Such behavior is captured by models with and without an energy cost associated with the interfaces, as the temperature-quenched Ising (IM0) and the Voter~\cite{HoLi75,Redner19} models (VM), respectively. 
These two models are relevant limits of the system we study here.
They are equivalent in space dimension $d=1$, but 
present morphologically different domains as $d>1$, because of the  mechanism that drives the evolution of their interfaces, surface tension (curvature) for the IM0 and interfacial noise for the VM.
As a consequence of the surface noise, in $d=2$ the interfaces between the domains remain rough and, differently from the power-law time dependence in Ising-like models, the fraction of active sites (those with opposite opinions) in the VM has a much slower evolution~\cite{DoChChHi01}. In $d\ge 3$ surface noise interrupts coarsening in the Voter model. 

The VM describes, in a simplified way, the opinion dynamics towards an absorbing state — the consensus, where one belief prevails. In the original VM, spin variables are located in the vertices of a lattice. At each step a randomly chosen spin copies the state of one of its nearest neighbor, chosen at random. Many variations of the VM have been proposed~\cite{StTeSc08,CaFoLo09},  for example, including long-range interactions~\cite{Corberi_20241d,corsmal2023ordering, corberi2024coarsening,corberi2024aging,Corberi_2024} or intermediate states between the original $\pm 1$ opinions~\cite{VoRe12,VeVa18,VaKrRe03,LaSaBl09,WaLiWaZhWa14,BaPiSe15,SvSw15}.
These states could represent a complete indecision (state 0) or more nuanced, less extreme positions (even continuous possibilities).
These intermediate states, along with other mechanisms that delay the transitions~\cite{StTeSc08,StTeSc08b,PeKhRo20,ViLlToAn24}, introduce some inertia in the process of switching between opposite options. 
A common feature among several of these models is an emergent surface tension, bringing these systems closer to the IM0 and leading observables to the consensus state through power-laws. 
Furthermore, the exponents associated with these power-laws are consistent with, but not exactly equal to, those observed in the single-flip dynamics of the IM0.

The VM can be exactly solved in any dimension or lattice~\cite{FrKr96,BeFrKr96}, and for any form of the interaction~\cite{Corberi_20241d,corsmal2023ordering, corberi2024coarsening,corberi2024aging,Corberi_2024}, due to the linearity of the equations governing the moments of its probability distribution. 
On the other hand, the non-Markovian variations with intermediate states are more difficult. 
At best, a mapping can be made onto a general field theory that justifies the similarities between these models and the IM0~\cite{DaGa08}.
Thus, most results in this class of systems is obtained through numerical simulations. 

Here we consider the Persistent Voter Model (PVM), introduced in Refs.~\cite{LaDaAr22,LaDaAr24}, where each opinion has two levels of confidence, extreme (zealots) and normal (regular voters). 
Zealots become resistant to state changes  regardless of the influence of their neighbors~\cite{Mobilia03,MoPeRe07,GaJa07,MaGa13,CoCa16}. 
This condition may be permanent for radical agents or, as is the case here, it is transient as the confidence may be reset, turning zealots into normal voters.
Transient zealots have more consistent opinions that do not change easily, akin to skepticism~\cite{AmAr18,AmDaAr20}.
Remarkably, the introduction of such inertial behavior usually optimizes the time to attain consensus.
The more general version of the PVM has non-Markovian transition rules, imposing a type of memory on each agent that determines whether or not they behave as zealots.
We propose here a simplified, Markovian version, showing that the memory mechanism does not necessarily need to be present for the most interesting properties of the model to be observed. 
Furthermore, being Markovian, an analytical, albeit approximate, approach becomes feasible. 
In particular, focusing on $d=1$ and $d=2$, we show that the power-law behavior of the density of active sites has the same exponent of model A universality class, as the IM0.

The paper is organized as follows. 
In the next section, the simplified version of the PVM is introduced and explained.
Section~\ref{sec.correlators} defines the dynamics of the model and the laws ruling the evolution of the correlators that describe it.
An approximate solution that closes the set of equations for the correlators at short distances is discussed in Sec.~\ref{sec.stat}.
Section~\ref{sec.largerr} discusses some approximations in order to evaluate the correlations at larger distances.
In all cases, results are obtained in both 1D and 2D and compared with extensive numerical simulations.
Finally, Sec.~\ref{sec.conclusions} summarizes and discusses the main results.

\section{The Persistent Voter model}
\label{sec.model}

In the standard Voter model (VM), a chosen agent compares its opinion (the binary variable $S_i=\pm 1$) with that of a nearest neighbor ($S_j$) and, if different, aligns with it: $S_i\to S_j$.
The Persistent Voter model (PVM)~\cite{LaDaAr22,LaDaAr24} introduces inertia into this process and an agent, if too confident in its opinion (a zealot), does not change its state.
The PVM is non-markovian since it may take several steps for an agent to increase its confidence $\eta_i$ and become a zealot (when attaining the threshold $\eta_i\geq 1)$.
Due to the positive reinforcement when interacting with an agent with the same opinion, the confidences $\eta_i$ and $\eta_j$ of both agents increase by $\Delta\eta$, a continuous, positive parameter.

\begin{figure}[ht]
    \begin{center}
    \includegraphics[scale=0.13]{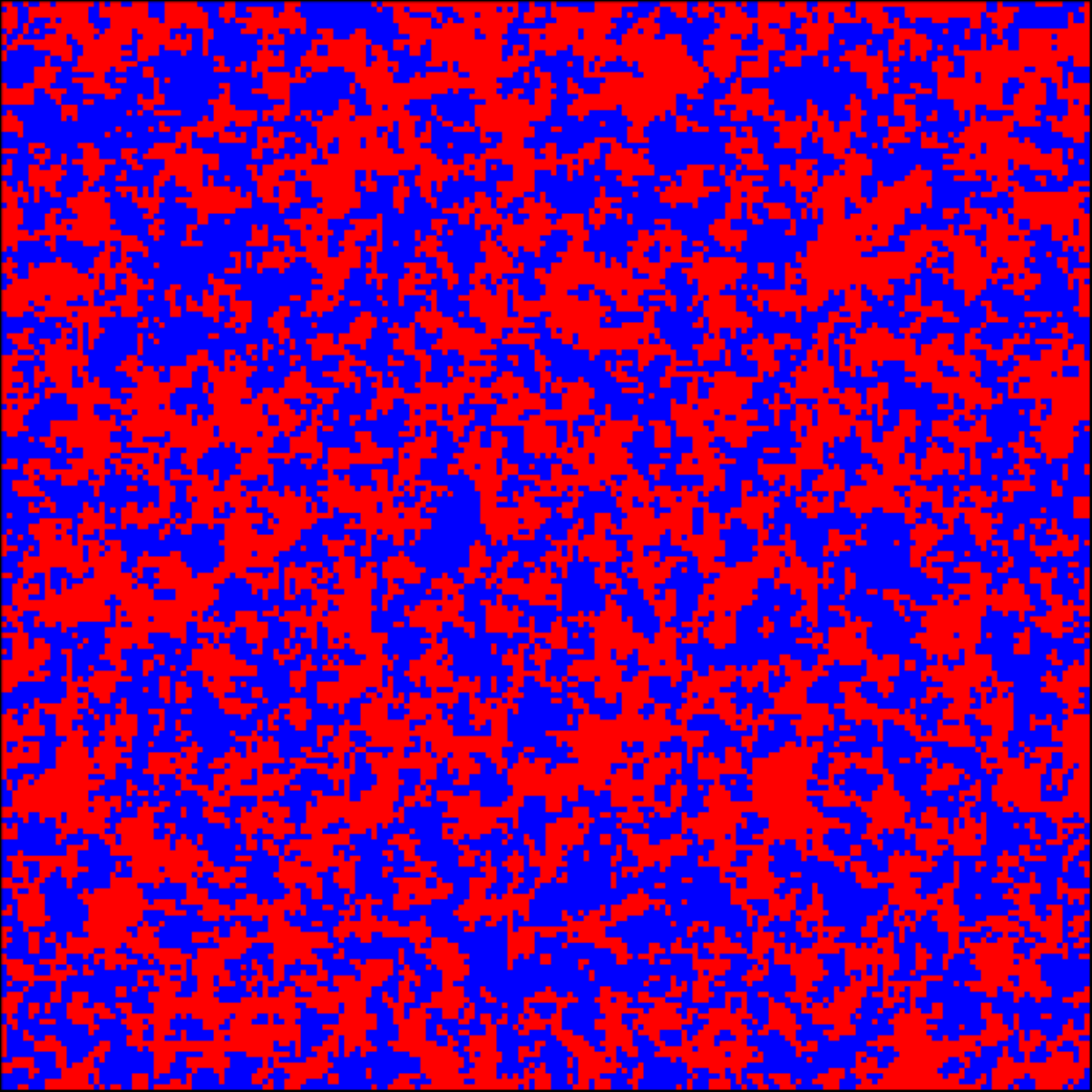}
    \includegraphics[scale=0.13]{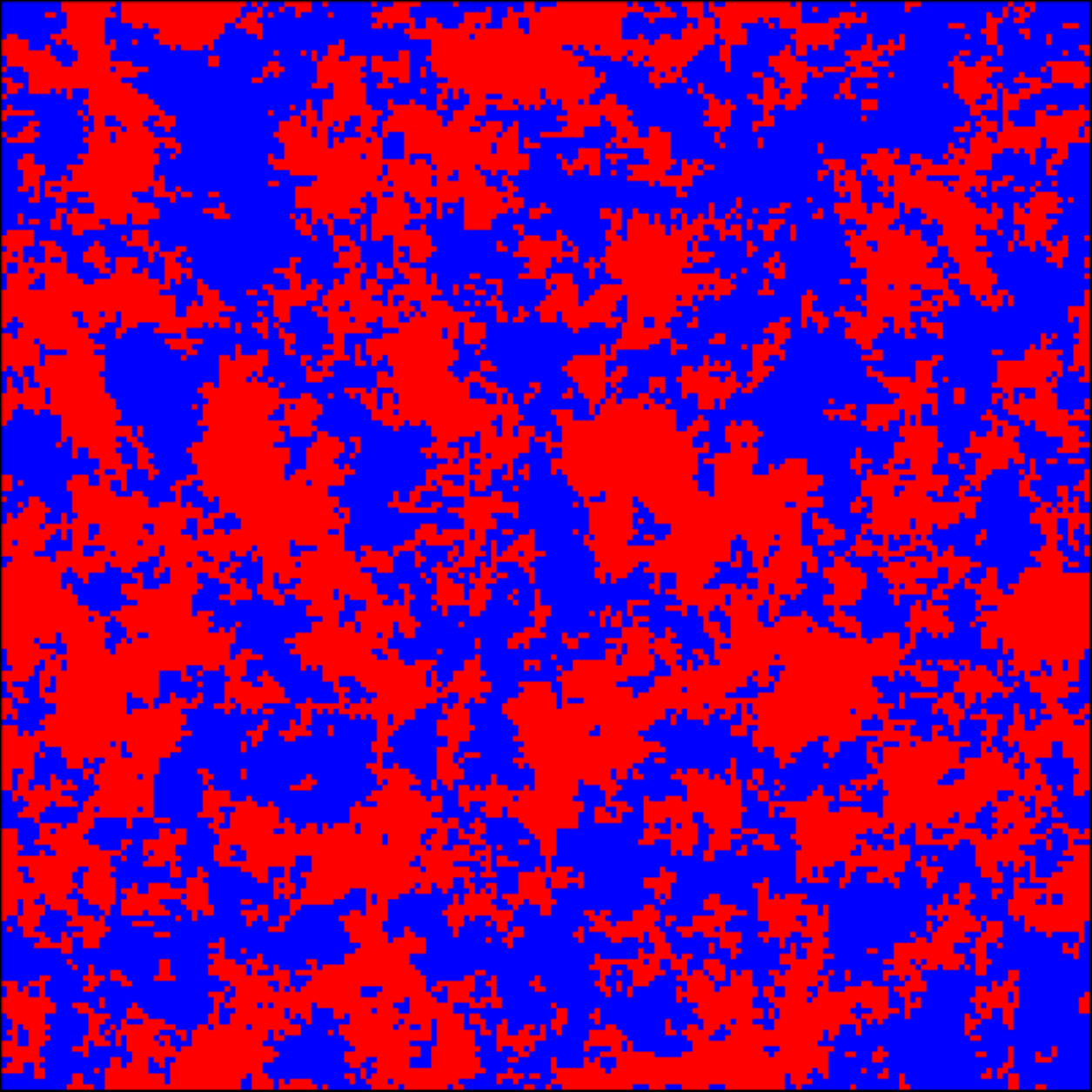}
    \includegraphics[scale=0.13]{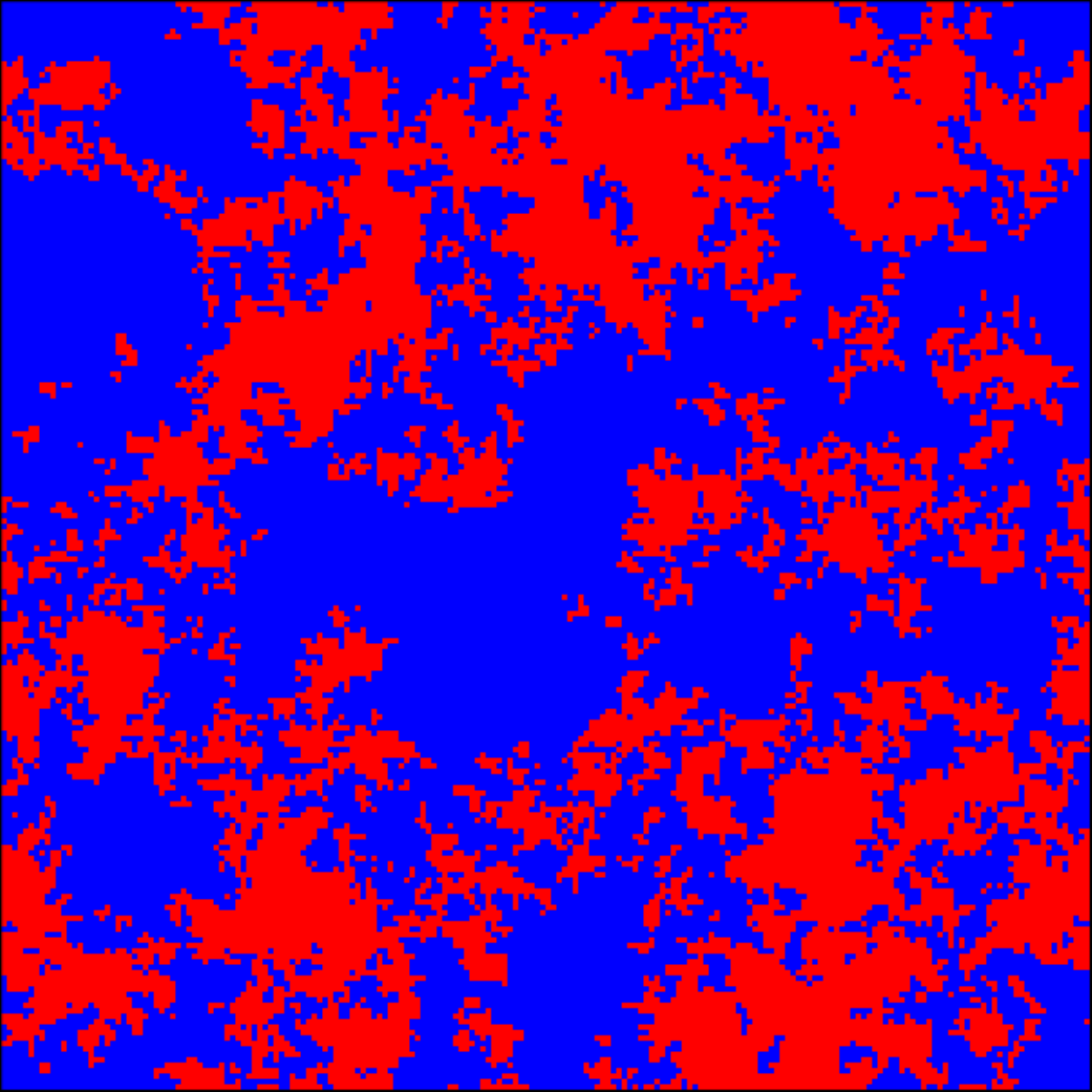}

    \includegraphics[scale=0.13]{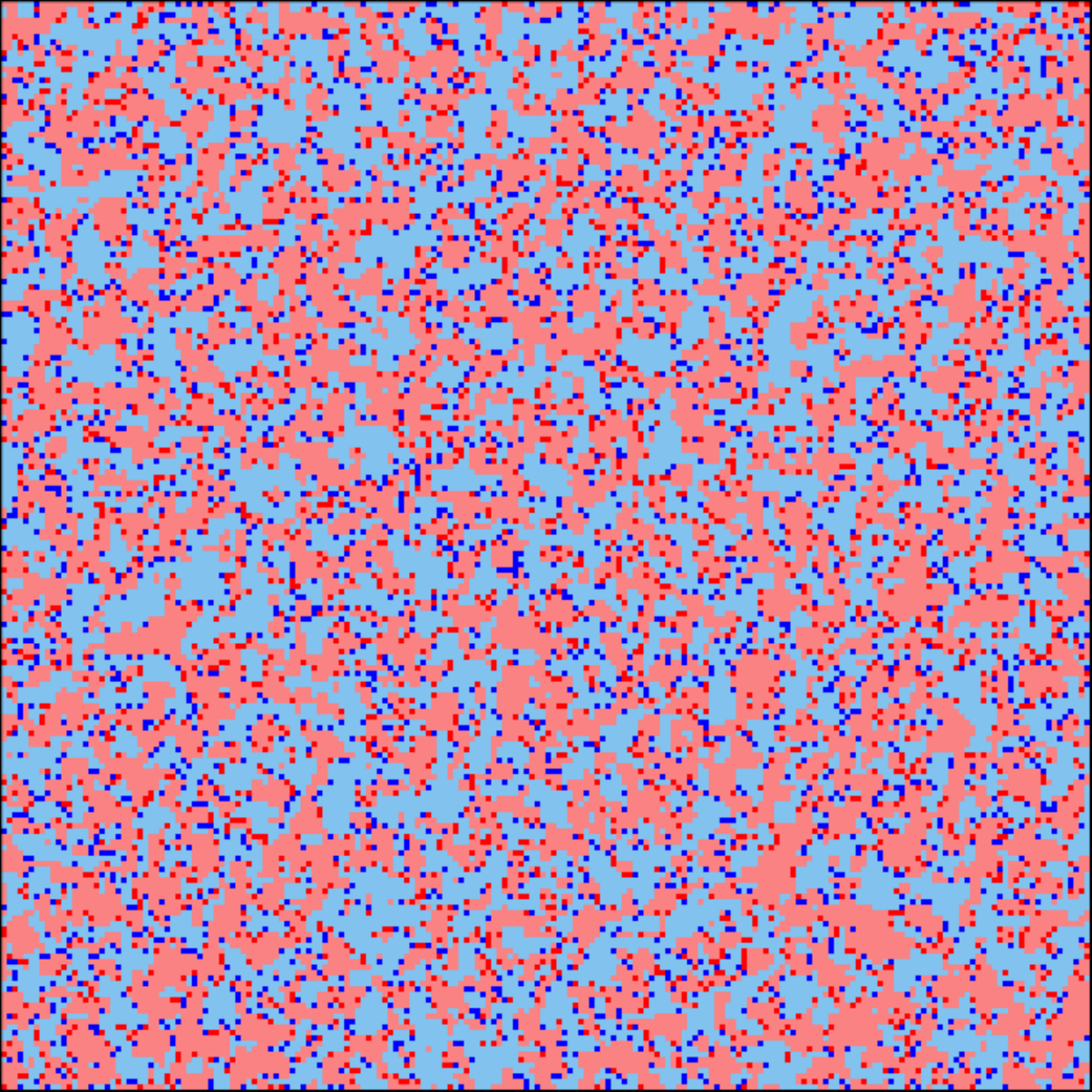}
    \includegraphics[scale=0.13]{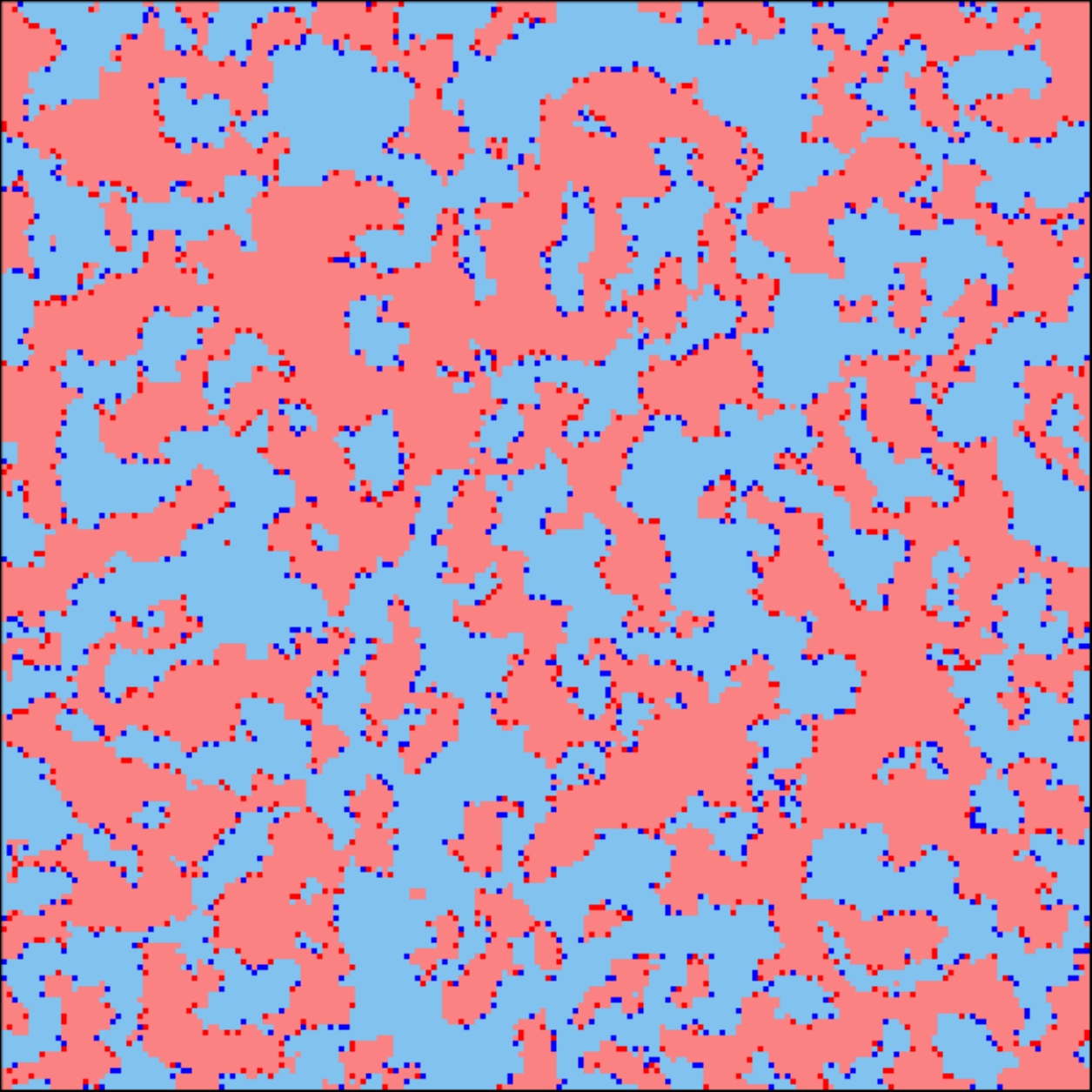}
    \includegraphics[scale=0.13]{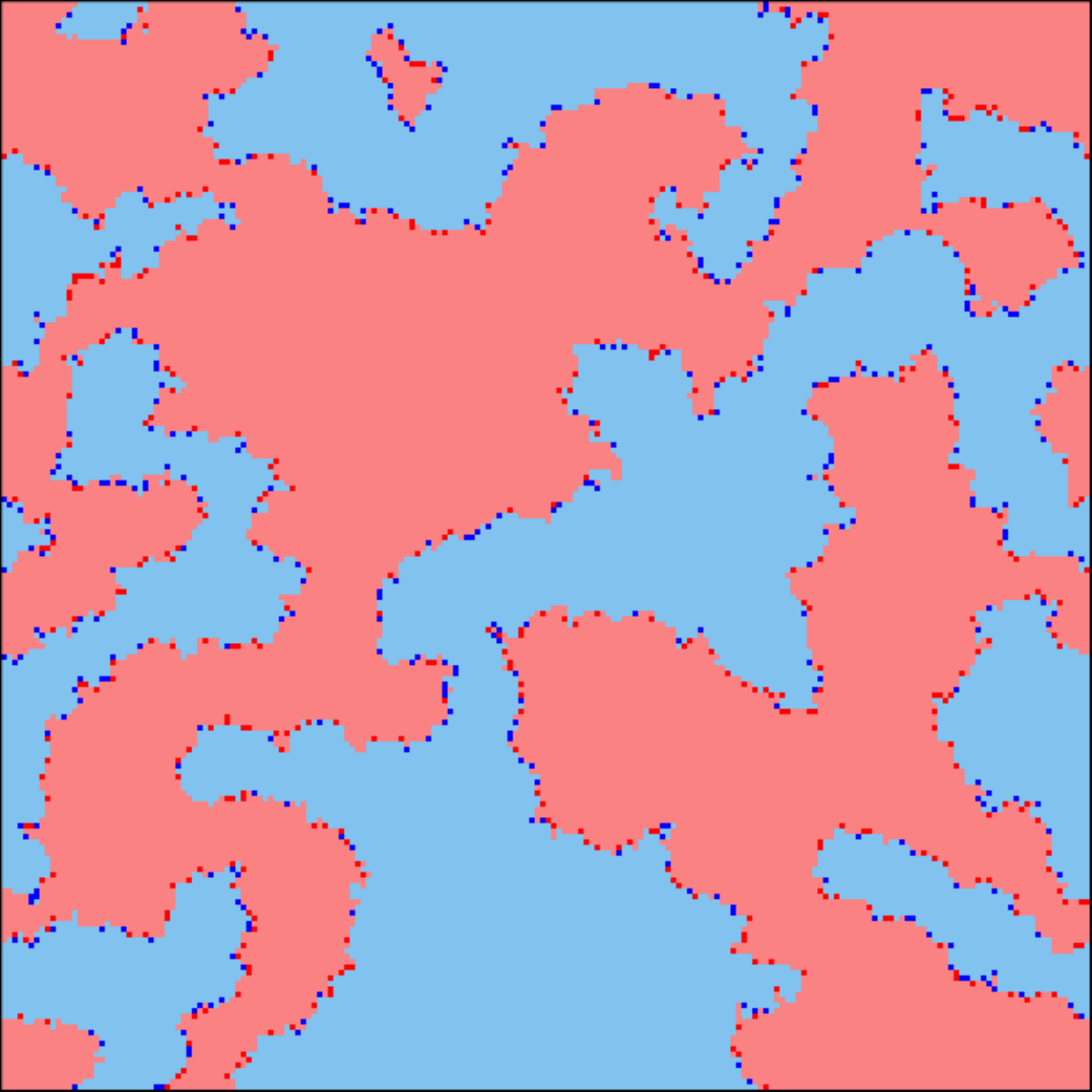}

    \includegraphics[scale=0.13]{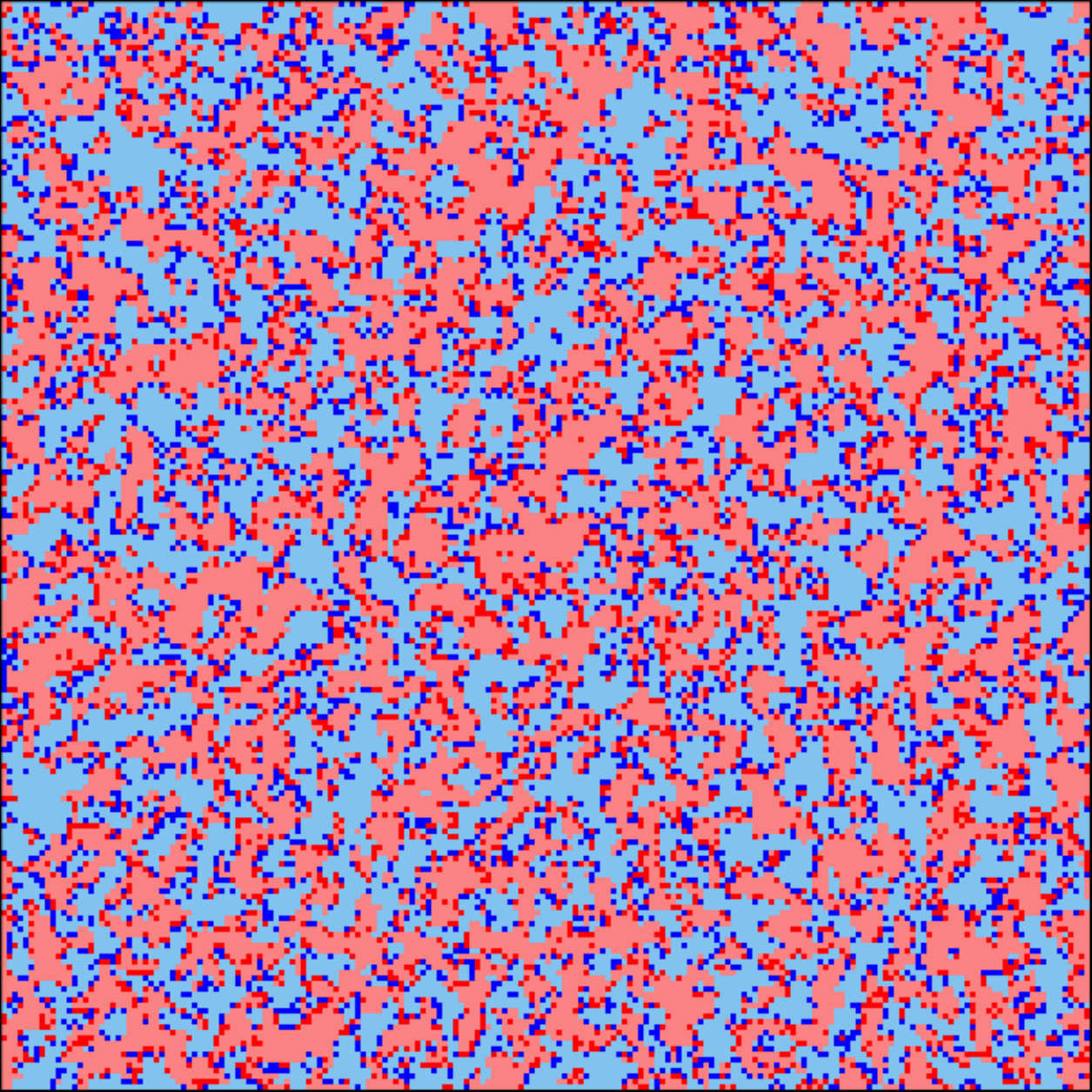}
    \includegraphics[scale=0.13]{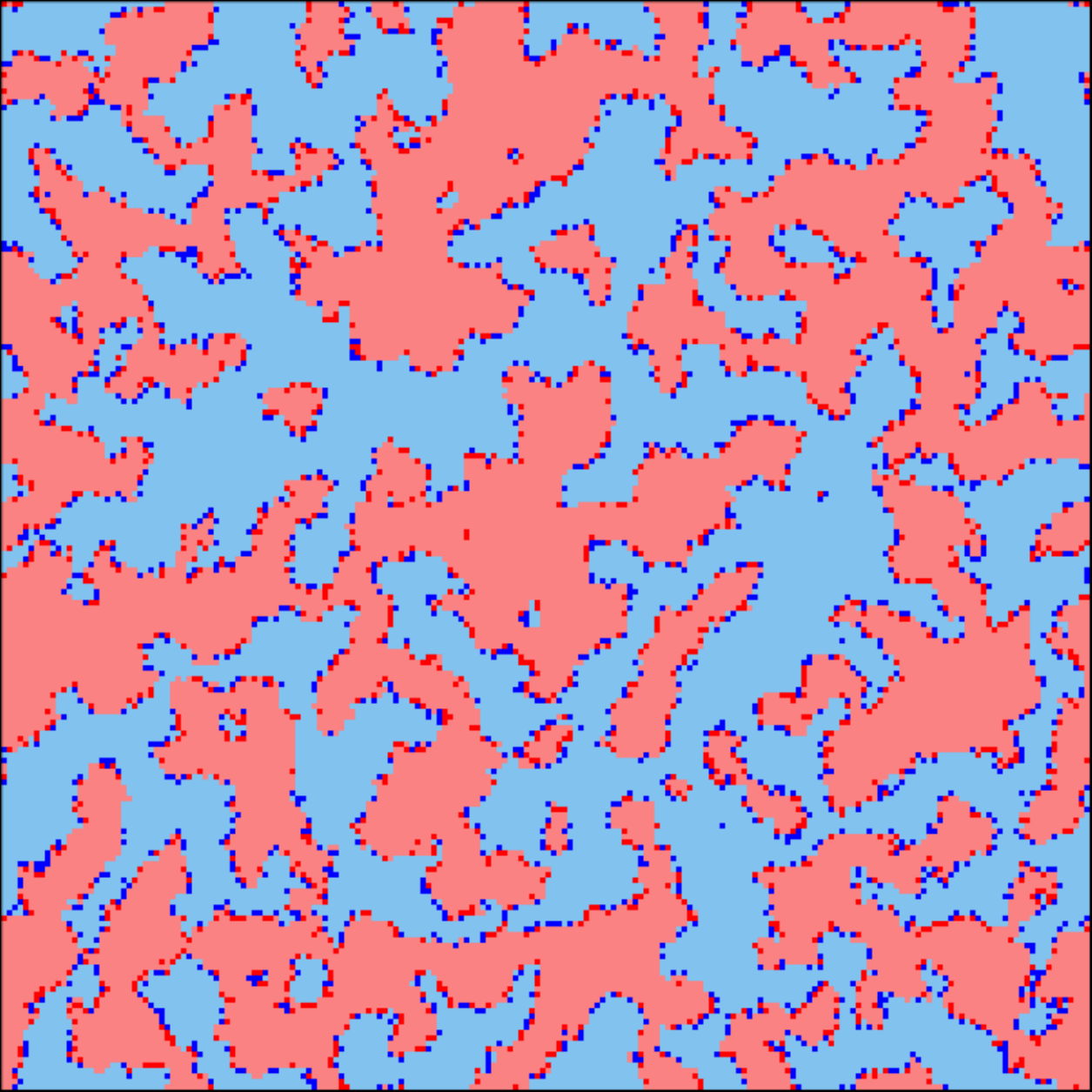}
    \includegraphics[scale=0.13]{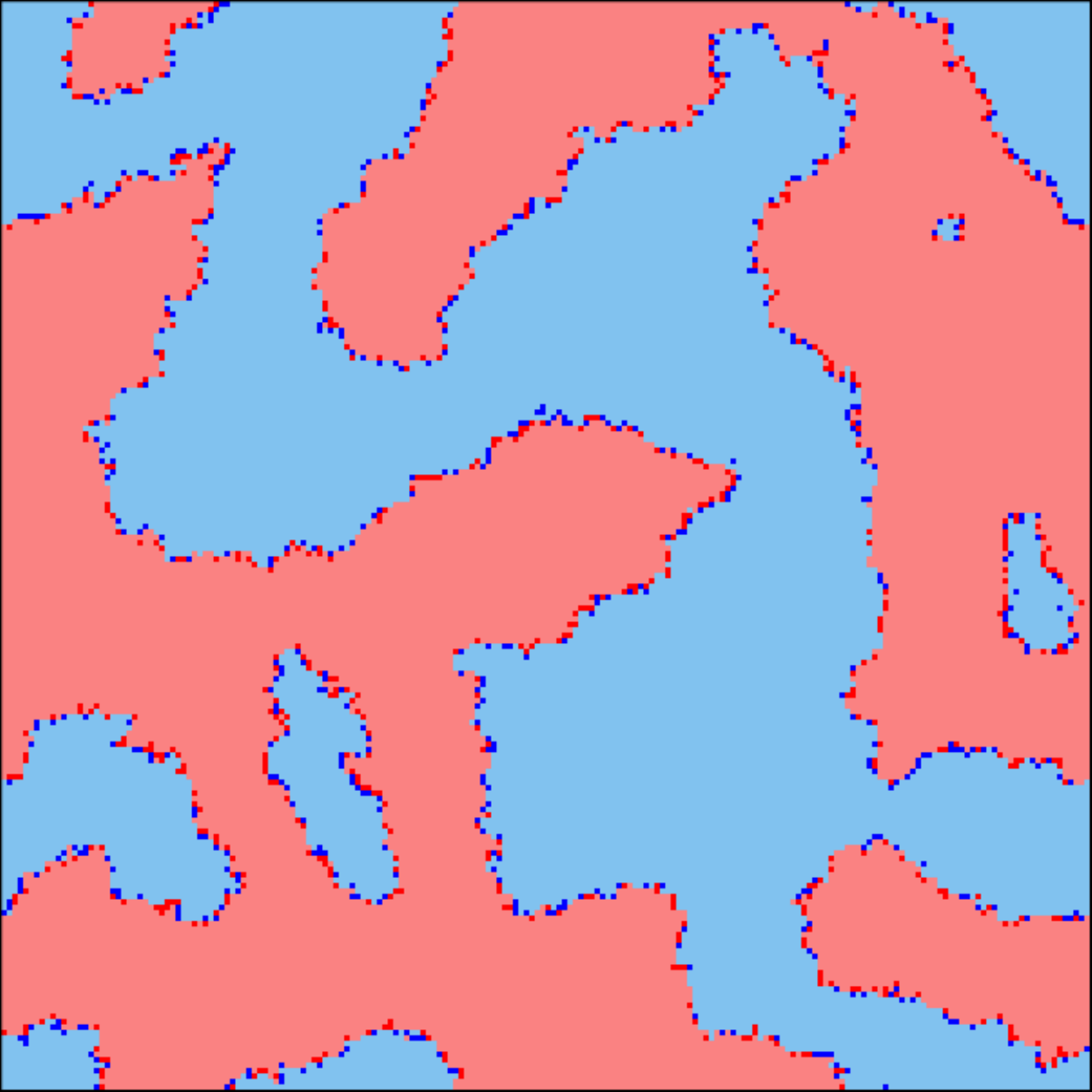}
    \caption{Snapshots for the VM (top row) where zealots are absent (dark colors are normal voters while light ones are zealots). The middle and bottom rows show the original and the simplified PVM, respectively, both with  $\Delta\eta=1$. In all cases, the snapshots are at $t=10$, 100 and 1000 MCS (from left to right) on a 2D lattice with length side $L=200$.}
    \label{fig.snapshots}
    \end{center}
\end{figure}

We consider a version of the PVM that, despite its simplifications, still captures the essential properties  of the original model, in particular the emergent coarsening behavior.
For $\Delta\eta=1$, the model becomes Markovian and, instead of the internal confidence, it is more convenient to use a second binary variable $\theta_i=+1$ for a zealot or $-1$ for a normal voter.
In this case, a normal voter may turn into a zealot upon a single interaction with an equally-minded neighbor (if $S_i=S_j$ then $\theta_i=1$) and, likewise, reverse to a normal voter after interacting with an agent with a different opinion (if $S_i\neq S_j$ then $\theta_i=-1$).
Moreover, in the original version, the confidence of the neighbor $j$ also increased in each interaction with the focal agent $i$ \cite{LaDaAr22,LaDaAr24}.
This seems not to be essential and will no longer be considered.
Fig.~\ref{fig.snapshots} shows snapshots of a 2D system with the spatial distribution of different opinions during the time evolution after starting with equally distributed random opinions for the VM (top row), the original (middle row) and modified (bottom row) PVM.
In the VM the domain growth is driven by the interface noise and the interfaces are not smooth.
However, in both versions of the PVM, zealots form in the bulk of the domains while normal voters concentrate close to the interfaces, forming a very thin layer.
These ingredients, combined, are responsible for the observed curvature-driven growth~\cite{LaDaAr22,LaDaAr24}.

\section{Correlators: definitions and temporal evolution}
\label{sec.correlators}

With the inclusion of zealots, the transitions for the opinion and zealot variables ($S_i\to -S_i$ and $\theta_i\to -\theta_i$ respectively) have rates given by
\begin{align}
  \label{eq.wsi}
  w(S_i)&=\frac{1}{2}\left(\frac{1-\theta_i}{2}\right)\left(1-\frac{S_i}{2d}\sum_{\delta}S_{i+\delta}\right)\\
  \label{eq.wti}
  w(\theta_i)&=\frac{1}{2}\left(1-\frac{S_i\theta_i}{2d}\sum_{\delta}S_{i+\delta}\right),
\end{align}
where the sum is over all nearest neighbors of $i$.
The time evolution of the correlator $\langle\alpha_1\alpha_2 \cdots \alpha_m\rangle$, where $\alpha_i$ may be either $S_i$ or $\theta_i$, is given by
\begin{equation}
  \label{eq.evcor}
  \frac{d}{dt} \langle\alpha_1\alpha_2\cdots\alpha_m\rangle=-2\left\langle\alpha_1\alpha_2\cdots\alpha_m\sum_{i=1}^m w(\alpha_i)\right\rangle. 
\end{equation}
To simplify the notation, we define 
\begin{equation}
  \label{eq.repcor}
  C_{i,j,k,...}^{n,p,q,...}\equiv\langle S_i S_j S_k\cdots\theta_n\theta_p\theta_q\cdots\rangle.
 \end{equation}
where subscript and superscript indices are for the $S$ and $\theta$ components, respectively. In
Appendix~\ref{ap.invariance} several useful
invariance symmetries of these correlators are listed as, for example,  $C_i=C_j$, $C^i=C^j$, 
$C_{i,j}= C_{i+r,j+r}$, etc, $\forall i,j$.
Some of the low order correlators, $m=1$ and 2 in Eq.~(\ref{eq.evcor}), that are useful in the following (there are other two-point correlations but they are not necessary for closing these equations) are 
\begin{align*}
    \frac{dC_i}{dt}=&\frac{1}{2}\left(C_i^i - C_i\right) +\frac{1}{4d}\sum_{\delta}C_{i+\delta}-\frac{1}{4d}\sum_{\delta}C_{i+\delta}^i \\
    \frac{dC^i}{dt}=&-C^i+\frac{1}{2d}\sum_{\delta}C_{i,i+\delta}\\
    \frac{dC_{i,j}}{dt}=& -C_{i,j}  +\frac{1}{4d}\sum_{\delta}C_{i+\delta,j} +\frac{1}{4d}\sum_{\delta}C_{i,j+\delta}\nonumber\\
    &+ C_{i,j}^i -\frac{1}{2d}\sum_{\delta}C_{i+\delta,j}^i.
\end{align*}

Unlike magnetization, which should be 0 by symmetry at least for the random initial conditions considered here, the equation for $C^i$ does not have a definite parity under the transformation $\theta_i \to -\theta_i$, which makes it non trivial. Indeed, as already commented regarding Fig.~\ref{fig.snapshots}, the number of non-zealots is related to the density of interfaces.
The above equations can be simplified by considering the translational
invariance symmetries (see Appendix~\ref{ap.invariance}):
\begin{align}
  \label{eq.cis}
  \frac{dC_i}{dt}&=\frac{1}{2}\left( C_i^i - C_{i+\delta}^i \right)\\
  \label{eq.cit}
  \frac{dC^i}{dt}&=-C^i+C_{i,i+\delta}\\
  \label{eq.cij}
    \frac{dC_{i,j}}{dt}&=-C_{i,j}+\frac{1}{2d}\sum_{\delta}C_{i+\delta,j}+C_{i,j}^i-\frac{1}{2d}\sum_{\delta}C_{i+\delta,j}^i,
\end{align}
where we kept the sums in order to distinguish between the cases in which $i$ and $j$ are nearest neighbors or not. 
In the former case, if $i$ and $j$ are nearest neighbors, using that $C_{i,i}=1$ and $C_{j,j}^i=C^i$, Eq.~(\ref{eq.cij}) becomes
\begin{align}
  \frac{dC_{i,i+\delta}}{dt} =& -C_{i,i+\delta}+\frac{1}{2d}+\frac{2d-1}{2d}C_{i,i+2\delta}+C_{i,i+\delta}^i+\nonumber\\
  &-\frac{1}{2d}C^i-\frac{2d-1}{2d}C_{i,i+2\delta}^i,
  \label{eq.cijneigh}
\end{align}
where $C_{i,i+2\delta}$ is the correlator for next-nearest neighbors sites. 

Because of the dependence on higher-order correlators, Eqs.~(\ref{eq.cis})-(\ref{eq.cij}) do not form a closed set. 
In order to close and solve them, we must resort to some approximation scheme. 
Next section presents and discusses the results obtained using one of such approximations.

\section{Statistically independent variables}
\label{sec.stat}

The simplest approximation is to assume that the spin and zealot variables are statistically independent. For example, $C_i^j  = C_iC^j$, $C_{i,j}^i=C^iC_{i,j}$, etc.
This assumption, despite being very crude, along with the closure approximation discussed below, allows for interesting results that remarkably agree with the simulations.
Within this approximation, Eq.~(\ref{eq.cis}) shows that the magnetization $C_i$ is conserved, 
\begin{equation*}
\frac{dC_i}{dt}=\frac{1}{2}\left(C_iC^i-C_{i+\delta}C^i\right)=0,
\end{equation*}
independently of any symmetry argument and in agreement with the simulations using random initial conditions.
Nonetheless, this conservation is not guaranteed in other scenarios, as the zealots may disrupt it.

\begin{figure}[htb]
\includegraphics[width=\columnwidth]{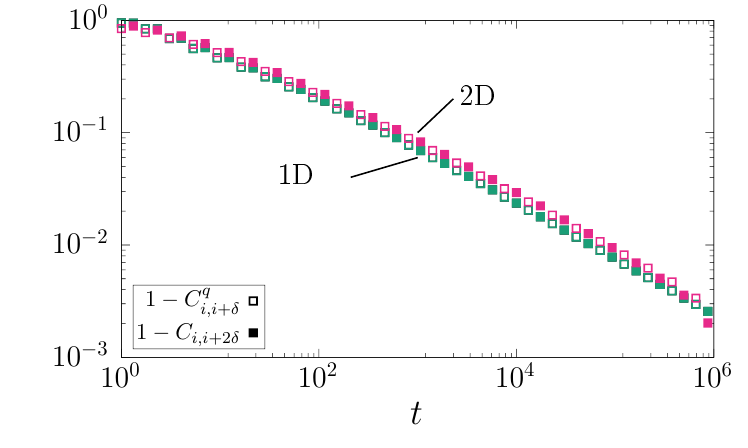}
\caption{Numerical test of the proposed {\it ansatz} for the closure of the set of equations for the low-order correlators. The correlator  for next-nearest-neighbors $C_{i,i+2\delta}$ (filled symbols) agrees very well with $C_{i,i+\delta}^q$ (hollow symbols), for both 1D ($q=2$) and 2D ($q=4/3$). The lattices have  $L=10^4$ and $L=10^3$, respectively, and averages are over $10^3$ samples with random initial conditions.}
\label{fig.nnn}
\end{figure}

The above assumption is not enough to close the remaining equations, (\ref{eq.cit}) and (\ref{eq.cijneigh}), and some further approximation must be considered.  
Notice that Eq.~(\ref{eq.cit}) remains the same while Eq.~(\ref{eq.cijneigh}) becomes
\begin{align}
    \frac{dC_{i,i+\delta}}{dt}
    =(1-C^i)\left[\frac{1}{2d}-C_{i,i+\delta}+\left(1-\frac{1}{2d}\right)C_{i,i+2\delta}\right].
    \label{eq.cii+1}
\end{align}
This shows that a possible stationary state is the consensus, $C_{i,i+\delta}=C^i=1$, where all agents share the same opinion and behave as zealots.
To close these equations it is necessary to express the next-nearest neighbor correlator, $C_{i,i+2\delta}$, as a function of the other variables. 
Supported by the numerical simulations shown in Fig.~\ref{fig.nnn}, we propose the following {\it ansatz}:
$C_{i,i+2\delta}\approx C_{i,i+\delta}^q$, where $q$ is a positive real number. 
The agreement is very good for both the 1D and 2D cases with $q=2$ and $4/3$, respectively (see below for an explanation of these values).

\begin{figure}[htb]
\includegraphics[width=\columnwidth]{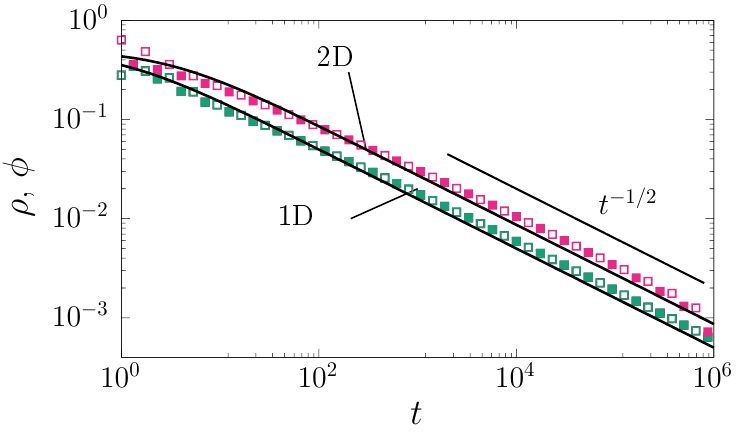}
\caption{Density of active sites $\rho$ (filled symbols) and the fraction of zealots $\phi$ (hollow symbols), in both 1D and 2D. At the early steps of the dynamics, both quantities become already very close, $\rho\simeq\phi$. Also, these quantities are well described by Eq.~(\ref{eq.rhod}), solid lines, and decay as $t^{-1/2}$ toward the consensus state ($\phi=\rho=0$). Same parameters of Fig.~\ref{fig.nnn}.}
\label{fig.interf}
\end{figure}

Defining the variables $\rho\equiv (1-C_{i,i+\delta})/2$, the density of active sites, and $\phi\equiv (1- C^i)/2$, the density of non-zealots, Eqs.~(\ref{eq.cit}) and (\ref{eq.cii+1}) can be rewritten as
\begin{align}
  \label{eqphi}
  \frac{d\phi}{dt}&=\rho-\phi\\
  \label{eqrho}
  \frac{d\rho}{dt}&=\frac{\phi}{2d}\bigg[ 2d(1-2\rho) - 1 +(1-2d)(1-2\rho)^q\bigg].
\end{align}
The fixed point is the consensus state where $\rho=\phi=0$, and all agents are zealots bearing the same opinion.
However, for large times, the system attains the scaling regime in which the domains coarsen and both quantities are very close, $\rho\simeq\phi$, while decreasing toward 0 as $t^{-1/2}$, what is also seen in the numerical simulation for 1D and 2D (see Fig.~\ref{fig.interf}).
In the bottom row of Fig.~\ref{fig.snapshots} we observe that the non-zealots agents are located close to the interfaces between the domains
since the evolution of both $\rho$ and $\phi$ primarily occurs at the domain boundaries, where sites with opposing states coexist. 
These agents may either interact with same-opinion agents in the bulk or with opposite-opinion ones across the boundary, thus the probability of being a zealot is $1/2$, as it can also be readily evinced by looking at the transition rate Eq.~(\ref{eq.wti}). 
The same is valid for the agents in the opposite boundary, making the total number of non-zealots be close to the length of the interface and $\rho\simeq\phi$.
Then, using that $\rho\simeq\phi$ for $t\gg 1$, Eq.~(\ref{eqrho}) depends only on $\rho$,
and for the consensus to be a stable solution, the following condition must hold
\begin{equation}
  \label{eqq}
  q\leq\frac{2d}{2d-1}.
\end{equation}
This result is derived from a second-order analysis, as $\rho=0$ is
only marginally stable under a linear stability test.
Notice that the stability threshold corresponds to the values of $q$ used in Fig.~\ref{fig.nnn}: $q=2$ (1D) and 4/3 (2D). 

At timescales where $\phi\simeq\rho$, and using the value of $q$ at the edge of stability (the equality in the above equation), Eq.~(\ref{eqrho}) can be expanded to second order,
$$
\frac{d\rho}{dt}+\frac{2\rho^3}{2d-1} =0,
$$
whose solution, given the initial condition $\rho(0)=1/2$, is
\begin{equation}
    \label{eq.rhod}
  \rho(t)=\frac{1}{2}\sqrt{\frac{2d-1}{2d-1+t}}.
\end{equation}
The comparison between this solution and the numerical simulations on a square lattice is shown in Fig.~\ref{fig.interf}. 
Besides observing that indeed $\rho\simeq\phi$ both in 1D and 2D, there is also a very good agreement with the above analytical results. 
Asymptotically, $\rho(t)\sim t^{-1/2}$, the observed exponent in the coarsening regime. Let us recall that this is different from the VM in $d=2$, where $\rho$ decreases logarithmically~\cite{DoChChHi01}. 

\begin{figure}[htb] 
\includegraphics[width=\columnwidth]{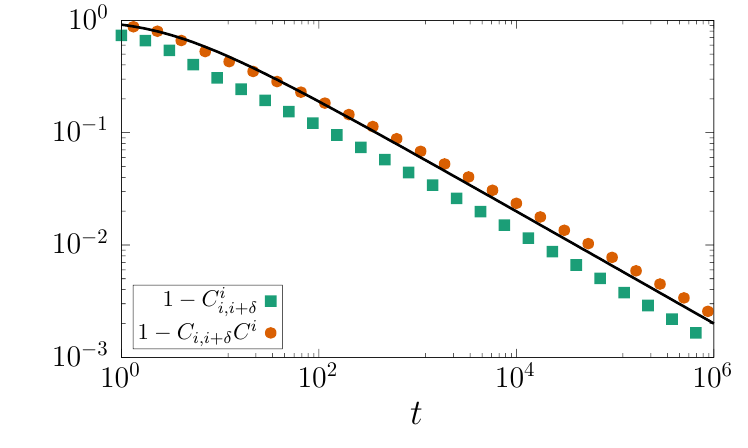} 
\caption{Comparison showing a significant difference between the correlator $C_{i,i+\delta}^i$ (squares), estimated from numerical simulations, and its approximate form, $C_{i,i+\delta}C^i\simeq 1-4\rho(1-\rho)$, with $\rho$ given  (solid line) by Eq.~(\ref{eq.rhod}) (we plot here the distance of these quantities from unity, for better visualization). The direct measure of $C_{i,i+\delta}C^i$ (circles), as previously shown, is well described by the implemented approximations. Numerical results are shown for a 1D system with $L=10^4$ but similar results are obtained in 2D.}    
    \label{corre3} 
\end{figure}

Despite Eqs.~(\ref{eq.cit}) and (\ref{eq.cijneigh})  providing a good description for $\rho$ and $\phi$ in the simulations, the assumption of statistical independence between $S$ and $\theta$ variables has its limitations, as shown in Fig.~\ref{corre3}. 
The correlator $C_{i,i+\delta}^i$, calculated from simulations, and $C_{i,i+\delta}C^i$, calculated from the solution of Eqs.~(\ref{eq.cit}) and (\ref{eq.cijneigh}), should be close if the approximation was correct. 
When comparing them, both in 1D and 2D (not shown), the approximate correlator underestimates the original one, as expected once the approximation neglects the correlation between the variables $S_i$ and $\theta_i$. Nonetheless, both present the same $~t^{-1/2}$ scaling at long times.

\section{Correlations at larger distances}
\label{sec.largerr}

The approximation discussed in the previous section gives reasonable results for correlations between nearest-neighbor agents. 
However, it fails when considering sites farther apart. 
To address this limitation, we propose an alternative method exploring the Laplacian operators in Eq.~(\ref{eq.cij}).
In addition, by connecting $q$ with the distance between sites in the correlator, we provide a further justification for the closure condition that was used in the previous section.

In Eq.~(\ref{eq.cij}), the r.h.s. represents the difference between two discrete Laplacians (see Appendix~\ref{ap.laplacian}):
\begin{align}
\frac{d}{dt} C_{i,j} &= \frac{1}{2 d} \left( \Delta C_{i,j} - \Delta C_{i,j}^i \right)\, ,
\label{eq.dCr}
\end{align}
where
\begin{align}
\Delta C_{i,j}  &= \sum_{\delta}C_{i+\delta,j} - 2 d C_{i,j} \nonumber \\
\Delta C_{i,j}^i &= \sum_{\delta}C_{i+\delta,j}^i - 2 d C_{i,j}^i.
\label{thelaplacians}
\end{align}
When zealots are absent, $\theta_i=-1$, $\forall i$, then $C_{j,k}^i=-C_{j,k}, \forall j,k$, and Eq.~(\ref{eq.dCr}) reduces to $dC_{i,j}/dt = d^{-1}\Delta C_{i,j}$, the equation for correlations in the Voter Model~\cite{FrKr96,BeFrKr96,KrReBe10}.
The primary challenge lies in establishing a reasonable relationship between $\Delta C_{i,j}$ and $\Delta C_{i,j}^i$, as will be discussed in the following sections.

\subsection{1D}

\begin{figure}[htb]
\includegraphics[width=\columnwidth]{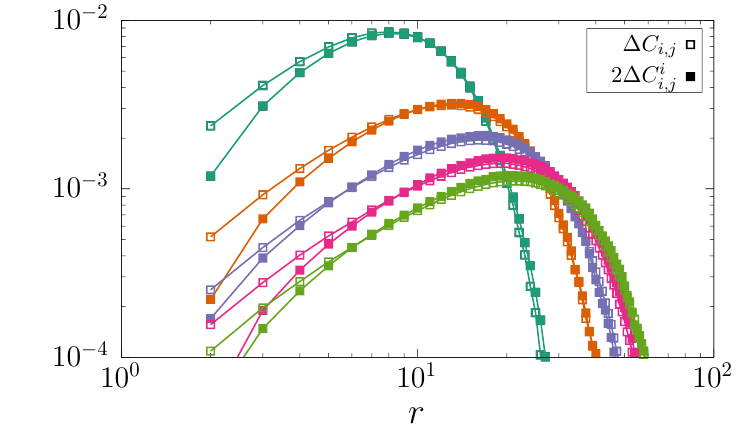} 
\caption{The laplacians defined in Eqs.~(\ref{thelaplacians}) computed in 1D are plotted versus $r\equiv |i-j|$ at different times ($t=100,300,\dots,900$ from top to bottom) indicating that the approximation $\Delta C_{i,j}^i\simeq\Delta C_{i,j}/2$ works well, except for very small $r$. 
System size is $N=10^5$ and averages are over $10^4$ samples.}
\label{comparison_laplacians}
\end{figure}

There are no contributions from the two Laplacians in Eq.~(\ref{eq.dCr}) in regions where the opinions around site $i$ are uniform, e.g., inside the bulk of a domain (at least one site away from an interface, as in the first two rows in Table~\ref{table.1d}).
All contributions originate from sites $i$ at the boundary (see Table~\ref{table.1d}).
When $\theta_i=+1$, the corresponding terms in the Laplacians of Eq.~(\ref{eq.dCr}) are equal and cancel out.
Because zealots and non-zealots appear with the same frequency along the boundary, half of those sites have $\theta _i=-1$ and contribute.
Hence, we assume that $\Delta C^i_{i,j}\simeq \Delta C_{i,j}/2$.
This approximation has been verified numerically, as shown in Fig.~\ref{comparison_laplacians}, and is indeed very accurate, except for small $r\equiv |i-j|$. This discrepancy can be understood because, when crossing very small domains (e.g., of unitary size), multiple  non-zealots may be encountered.
With this approximation, Eq.~(\ref{eq.dCr}) transforms into the usual diffusion equation of the Ising (or voter) model, but with an extra factor $1/2$.

\begin{table}[htb]
\begin{tabularx}{0.98\columnwidth}{|c|>{\centering\arraybackslash}X|>{\centering\arraybackslash}X|>{\centering\arraybackslash}X|>{\centering\arraybackslash}X|}
\hline
\includegraphics[width=3cm,trim= 0 8mm 0 0]{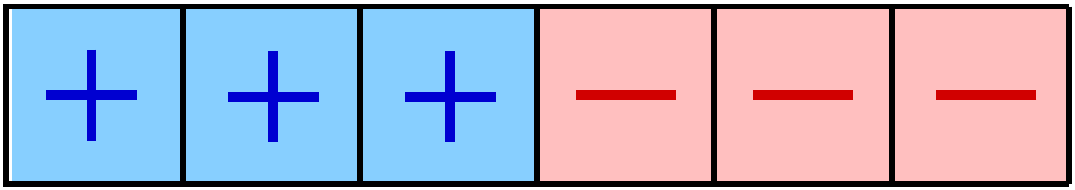}  & $C_{i-1,j}$ & $C_{i+1,j}$ & $C_{i,j}$ & $\Delta C_{i,j}$ \\
\hline
\includegraphics[width=3cm,trim= 0 8mm 0 0]{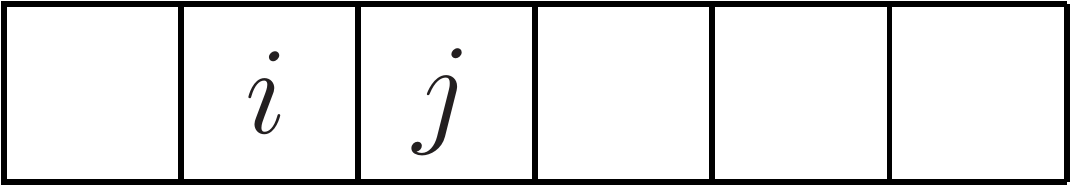} & $+1$ & $+1$ & $+1$ & 0   \\

\includegraphics[width=3cm,trim= 0 8mm 0 0]{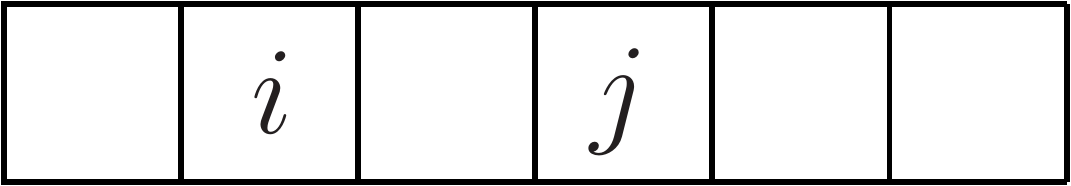} & $-1$ & $-1$ & $-1$ & 0 \\

\includegraphics[width=3cm,trim= 0 8mm 0 0]{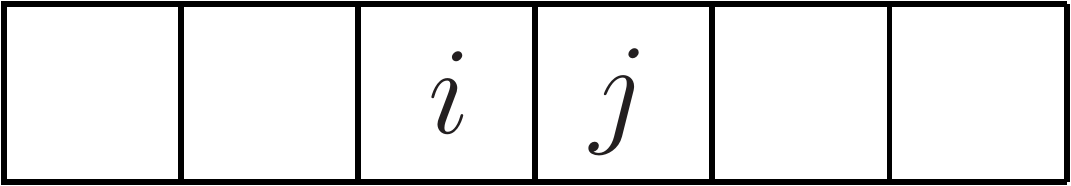} & $-1$ & $+1$ & $-1$ & $+2$  \\

\includegraphics[width=3cm,trim= 0 8mm 0 0]{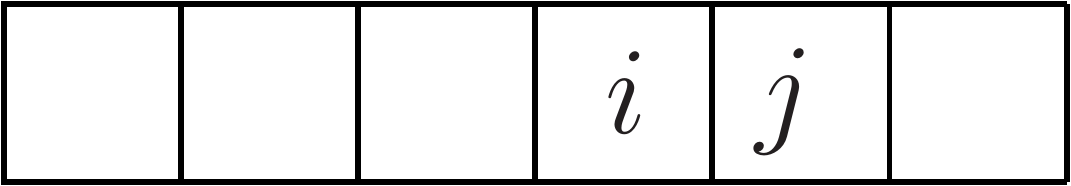} & $-1$ & $+1$ & $+1$ & $-2$ \\
\hline
\end{tabularx}
\caption{Examples of contributions to the Laplacians in Eq.~(\ref{thelaplacians}) for 1D, considering the domain wall between two large spin domains.  The first column depicts the interface  (different colors for up and down spins) and some examples of possible positions for $i$ and $j$ with $r=1$ and 2. Only the Laplacian $\Delta C_{i,j}$ is shown, in the last column.  
The evaluation of the second Laplacian, $\Delta C_{i,j}^i$, in this equation is similar but depends on whether the agent at site $i$ is a zealot or not (if a zealot, the two Laplacians cancel out in Eq.~(\ref{eq.dCr}), thus, only non-zealots contribute).
If site $i$ is away from the interface (first two rows), both Laplacians are null.}
\label{table.1d}
\end{table}

\begin{figure}[htb]
\includegraphics[width=\columnwidth]{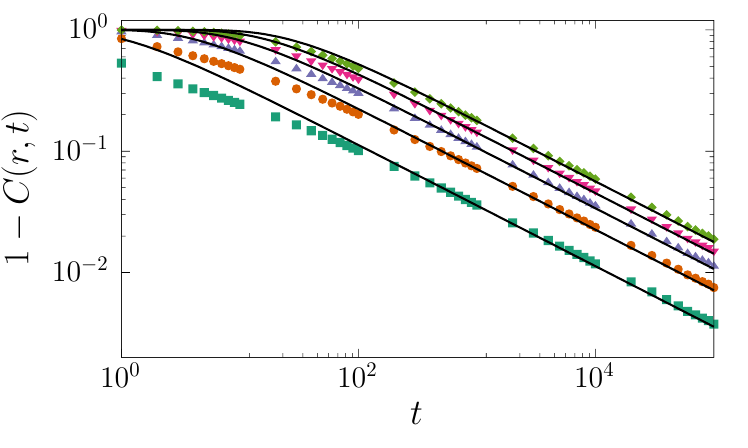}
\caption{Comparison between the 1D simulations of the PVM with $L=10^4$ (points) and  Eq.~(\ref{solcr1d}) (lines) for the correlator $C(r,t)$ and different values of $r$ (from bottom to top, $r=1, \ldots, 5$).}
    \label{long1d}
\end{figure}

For $r\gg 1$, because of the slow variation of $ C_{i,j}$ with $r$, we may use a continuous description  $C_{i,j} \simeq C(r,t)$  and the Laplacians in Eq.~(\ref{eq.dCr}), within this approximation, become
\begin{equation*}
  \frac{\partial}{\partial t}C(r,t)=\frac{1}{4}\frac{\partial^2}{\partial r^2} C(r,t).
\end{equation*}
If dynamical scaling holds, $r$ must scale with the dynamical correlation lenght $\ell(t)$. 
Using the ansatz $C(r,t)= f(r/\ell(t))$, we get
\begin{equation*}
-x \ell\dot \ell f'(x)=\frac{1}{4}f''(x),
\end{equation*}
where $x=r/\ell$.
In order to cancel the explicit time dependence, $\ell\dot{\ell}$ must be a constant and $\ell =D t^{1/2}$.
The above equation then simplifies
and its general solution is $f(x)=c_1+c_2\erf(Dx)$, 
where $\erf (x)$ is the error function.
Imposing the limits $C(r\to\infty,t)=0$ and $C(r,t\to\infty)=1$, which imply that $f(x\to\infty)=0$ and $f(0)=1$, one obtains $f(x) = \erfc(Dx)$, where $\erfc (x)$ is the complementary error function.
The correlator $C(r,t)$ then reads
\begin{equation}
  \label{solcr1d}
  C(r,t)=\erfc\left(\frac{r}{\sqrt{t}}\right),
\end{equation}
and its asymptotic behavior, for $t\gg 1$, is
\begin{equation}
  \label{asym1d}
  C(r,t)=1-\frac{2 r}{\sqrt{\pi t}}+\mathcal{O}\left(\frac{r^3}{t^{3/2}}\right).
\end{equation}
Figure~\ref{long1d} compares Eq.~(\ref{solcr1d})
with the numerical simulations for a 1D system with $L=10^4$. 
The agreement is very good and improves for $t\gg 1$ and larger values of $r$.

\subsection{2D}

\begin{figure}[ht!]
    \includegraphics[width=0.49\columnwidth]{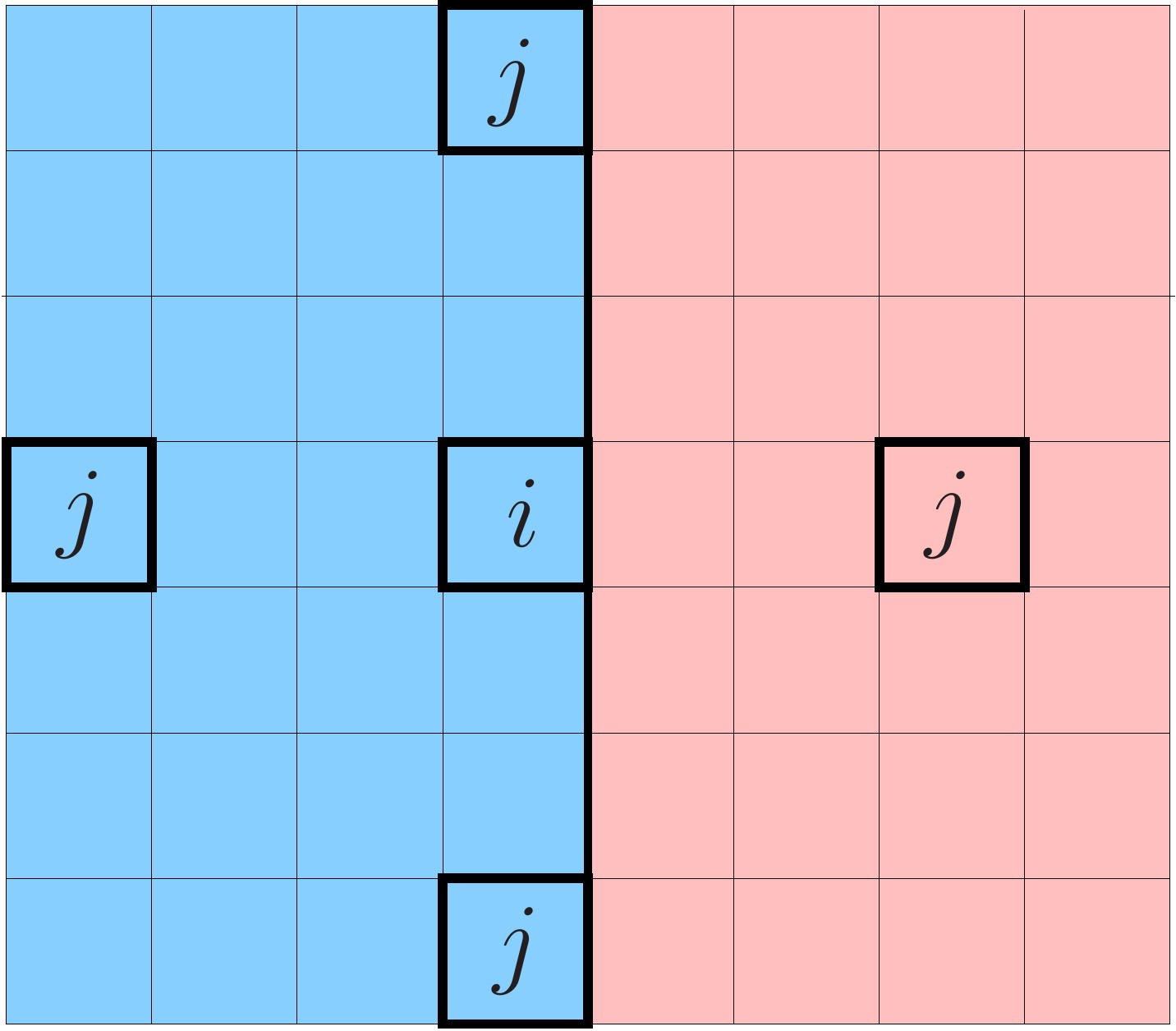}
    \caption{A vertical, flat interface between two domains with spins of opposite signs. Contributions to the Laplacians $\Delta C_{i,j}$ and $\Delta C_{i,j}^i$ come only from sites $i$ located along the interface. Sites $j$ are at a distance $r$ from $i$, as mentioned in the text.
    }
    \label{fig_interf_schem}
\end{figure}

\begin{figure}[htb]
\includegraphics[width=\columnwidth]{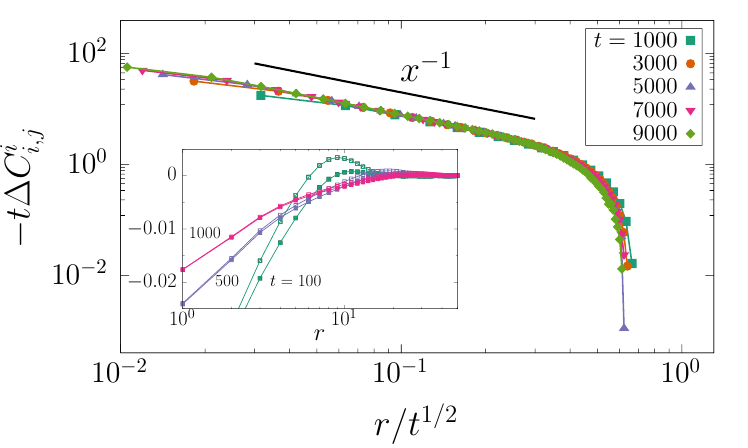}
\caption{Laplacian $\Delta C_{i,j}^i$ at different times showing an excellent collapse when $r$ is rescaled by $t^{1/2}$ (although the figure emphasizes the negative region of the Laplacians, the collapse is equally good in the whole measured range of $r$). The system size is $L=1000$ and averages are taken over 1000 samples. Inset: The two Laplacians at different times, $\Delta C_{i,j}$ (empty symbols) and $\Delta C_{i,j}^i$ (filled symbols). Although early in the dynamics ($t=100$, green points) there is still some difference, at larger times ($t=500$ and 1000) they become very close and the negative region, more prominent.}
    \label{fig.lap2d}
\end{figure}

Numerical simulations in 2D show that $\Delta C_{i,j}$ and $\Delta C_{i,j}^i$ have similar behaviors, albeit qualitatively different from the 1D case shown in Fig.~\ref{comparison_laplacians}.
Indeed, both Laplacians may become negative, as shown in the inset of Fig.~\ref{fig.lap2d} and, for large times, are almost equal.
In order to understand this result, consider a single, flat interface separating the two opinions, as schematically represented in Fig.~\ref{fig_interf_schem}.
In both Laplacians, site $i$ must be at the interface,  
otherwise there is no contribution due to the homogeneous neighborhood, as already discussed in 1D. 
Consider, as indicated in Fig.~\ref{fig_interf_schem}, $i$ at the interface and the sites $j$ located at a distance $r$ from $i$ along the horizontal and vertical directions.
The contributions to $\Delta C_{i,j}$ and $\Delta C_{i,j}^i$ coming from the two possible $j$ sites along the direction orthogonal to the interface (horizontal in the figure) have opposite signs and hence cancel each other in each of the Laplacians (the same conclusion applies to all other sites $j$ not at the interface since there is always a specular site relative to the column to which $i$ belongs).
Only the sites $j$ along the vertical interface (belonging to the same domain as $i$) sum up constructively their contributions, which turn out to be negative for $\Delta C_{i,j}$. As time increases, the surviving domains are large and have smooth interfaces, their contributions being similar to the flat case, making the above argument even more plausible. 

With this understanding, we can evaluate $\Delta C_{i,j}^i$
recalling where the most important contributions to this Laplacian come from: first, we need two sites 
$i$ and $j$ with parallel spins, which we can argue to happen with probability $C_{i,j}$. Furthermore, these sites must be located at the interface, which happens, roughly, with probability $\rho^2$. 
They must also belong to the same domain, what  occurs with an approximate probability proportional to $\ell/r$, where $\ell=\ell(t)$ is a measure of the average domain radius. 
Combining these informations, we have
$\Delta C^i_{i,j}\propto -\rho ^2C_{i,j}\ell/r$.
The minus sign appears because we know
that $\Delta C_{i,j}^i \simeq \Delta C_{i,j}$, the latter being negative. Besides being observed in the inset of Fig.~\ref{fig.lap2d}, this can be understood on the basis of geometry,
as discussed in Appendix~\ref{ap.equallaplacians}. 
Assuming scaling, one has $\ell\propto \rho^{-1}$, and then
\begin{equation}
    \Delta C_{i,j}^i\propto -\rho\, \frac{C_{i,j}}{r}
    \propto -(1-C_{i,i+\delta})\, \frac{C_{i,j}}{r}.
    \label{physical}
\end{equation}
Using, in this hypothesis, the property 
\begin{equation}
C_{i,j}=f\left [\frac{r}{\ell(t)}\right ]    
\label{scalcd2}
\end{equation}
(which is proved to be correct in Fig.~\ref{fig_approx2d_scalf}, and commented further on), one has the following scaling form 
for the Laplacian
\begin{equation}
\Delta C^i_{i,j}= -k \, \ell^{-2}(t) \,x^{-1} \, f(x) \, , \qquad x \equiv r/\ell(t) \, , 
\label{scaling2d}
\end{equation}
where $k$ is a positive proportionality constant, with the conditions
$x^{-1} f(x)\propto x^{-1}$ for $x\ll 1$, and $x^{-1} f(x) \to 0$ for $x\gg 1$. Eq.~(\ref{scaling2d}) is well verified numerically (with $\ell\propto t^{1/2}$), as can be seen in Fig.~\ref{fig.lap2d}.
Therefore, the actual form of $f$ found in simulations agrees with the overall behaviors reported above. Clearly, given the approximate nature of our approach, one can observe some quantitative differences between the scaling function found in simulations and its estimated form, Eq.~(\ref{scaling2d}).
For instance, the fast decrease of $\Delta C^i_{i,j}$ around $x=1$ in Eq.~(\ref{scaling2d}) is less sharp than in Fig.~\ref{fig.lap2d}. 

\begin{figure}[ht!]
    \begin{center}
    \includegraphics[width=\columnwidth]{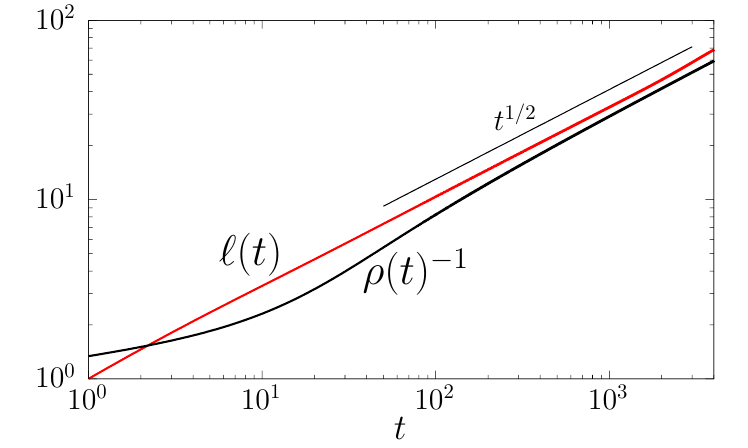}
    \caption{Numerical solution of Eq.~(\ref{eq.dCr}) in 2D with closure Eq.~(\ref{physical}), showing $\ell(t)$ and $\rho(t)^{-1}$.
    System size is $N=300^2$.}
    \label{fig_approx2d}
    \end{center}
\end{figure}

Plugging the above closure scheme, Eq.~\eqref{scaling2d} into Eq.~\eqref{eq.dCr}, using a continuous approach, and recalling that the Laplacian in polar coordinates is $\Delta C(r)= C''(r)+r^{-1} C'(r)$, we get
\begin{eqnarray}
  xf''(x)+\left(1+2 D^2x^2\right)f'(x)+kf(x)=0 \, , 
  \label{eqf}
\end{eqnarray}
where we took care of the explicit time-dependence by means of the ansatz $\ell= D t^{1/2}$.
We have tested this hypothesis by performing a numerical integration of Eq.~(\ref{eq.dCr}), with 
$\Delta C_{i,j}^i$ expressed as in Eq.~(\ref{physical}) (last expression), on a square lattice. In Fig.~\ref{fig_approx2d} we plot the quantities $\rho(t)$ and $\ell (t)$,
the latter being extracted from $C_{i,j}$ as 
\begin{equation}
    \ell(t)\propto \frac{\int r\,C(r)\,dr}{\int C(r)\,dr}\,.
\end{equation}
The assumed behavior $\rho(t)^{-1}\propto \ell(t)\propto t^{1/2}$ is thus confirmed.

Now we turn to determine the form of the scaling function $f(x)$.
Considering the asymptotic limit $x\ll 1$ (equivalent to $t\gg 1$), the term $x^2$ can be neglected in Eq.~(\ref{eqf}) and the solution of the simplified equation is
\begin{eqnarray}  f(x)=C_1J_0\left(2\sqrt{kx}\right)+C_2Y_0 \left(2\sqrt{kx}\right),
\end{eqnarray}
where $J_0(x)$ and $Y_0(x)$ are the Bessel functions of the first and second kind, respectively. By imposing the condition $f(0)=1$ and requiring that the solution be monotonically decreasing,
we find,
\begin{eqnarray}
  \label{solcr2d}
  C(r,t)=J_0\left(2\sqrt{\frac{\kappa r}{t^{1/2}}}~\right),
\end{eqnarray}
where $\kappa=k/D$, and its asymptotic behavior 
is given by
\begin{eqnarray}
  \label{asym2d}
  C(r,t)=1-\frac{\kappa r}{\sqrt{t}}+\mathcal{O}\left(\frac{r^2}{t}\right).
  \end{eqnarray}
The above solution was derived from the hypothesis $x \ll 1$. In the other extreme, the relevant equation becomes
\begin{eqnarray}
  f''(x)+2 D^2 x f'(x)=0 \, , 
\end{eqnarray}
which is equivalent to the 1D case. Therefore, its solution under the condition $f(x\to \infty)=0$ is
\begin{equation} 
C(r,t)=c\erfc\left(\frac{ r}{\sqrt{t}}\right) \, .
\label{2dlargex}
\end{equation}
Notice that although the constant $c$ could be set to 1 imposing $C(0)=1$, $x=0$ is beyond the range of validity of the present approximation.
%

Our analytical results are compared with the numerical solution of Eq.~(\ref{eq.dCr}), with the closure Eq.~(\ref{physical}), in Fig.~\ref{fig_approx2d_scalf}. Here we plot $C_{i,j}$, at different times, against $x=r/\ell(t)$, obtaining a perfect data collapse, indicating that 
Eq.~(\ref{scalcd2}) is obeyed. The numerically obtained scaling function agrees with the analytical forms Eqs.~(\ref{solcr2d}) and ({\ref{2dlargex}), for small and large values of $x$, respectively.

\begin{figure}[ht!]
    \begin{center}
    \includegraphics[width=\columnwidth]{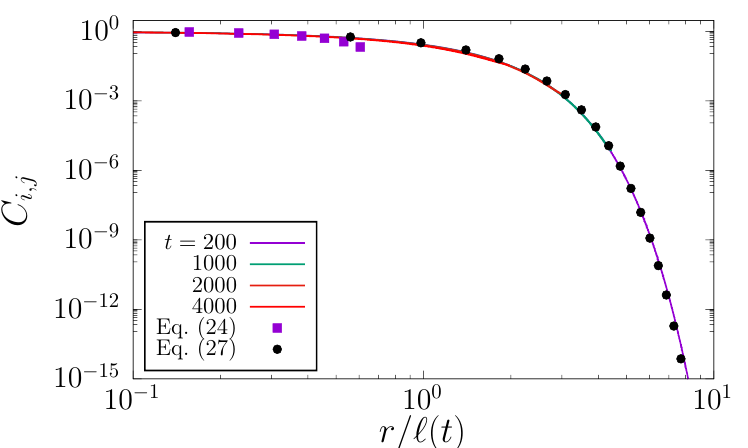}
    \caption{Numerical solution of Eq.~(\ref{eq.dCr}) for $C_{i,j}$, in 2D with closure Eq.~(\ref{physical}), plotted against $r/\ell(t)$ for different times (see key).
    System size is $N=300^2$. Purple squares are Eq.~(\ref{solcr2d}), while black circles are Eq.~(\ref{2dlargex}).}
    \label{fig_approx2d_scalf}
    \end{center}
\end{figure}

Now we compare our analytical findings directly with numerical simulations of the PVM. This is done in 
Fig.~\ref{long2d}, where simulation data for $C(r,t)$ are confronted with  the behavior predicted by Eq.~(\ref{solcr2d}).
The agreement is good and, similarly to the 1D system, it improves
in the long-time regime, consistent with the validity of the equation used to derive Eq.~(\ref{solcr2d}).

\begin{figure}[ht!]
    \includegraphics[width=\columnwidth]{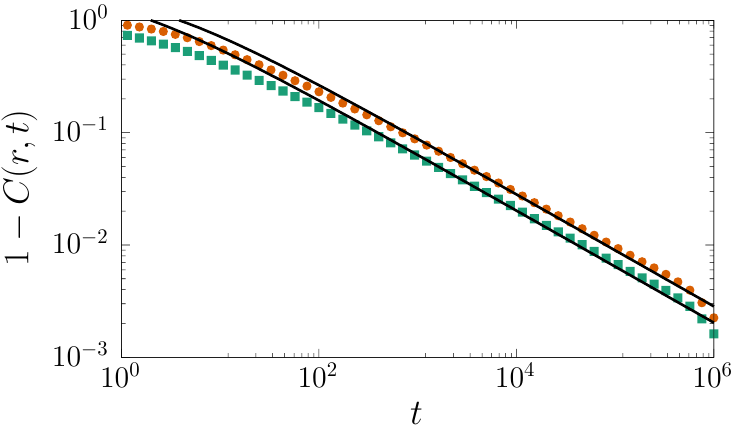}
    \caption{The 2D correlator $C(r,t)$, measured in a simulation with $L=1000$ and averages over 100 samples (symbols) and compared with Eq.~(\ref{solcr2d}). The parameter $\kappa\simeq 2.0$ is obtained through a fit with Eq.~(\ref{solcr2d}) and $10^3 \leq t\leq 10^5$. The bottom (top) curve corresponds to $r=1$ ($r=\sqrt{2}$).}
    \label{long2d}
\end{figure}
    
Finally, Eqs.~(\ref{asym1d}) and (\ref{asym2d}) can help to understand the motivation for the {\it ansatz} $C_{i,i+2\delta}\approx C_{i,i+\delta}^q$ used in Sec.~\ref{sec.stat}.
Both can be written as
\begin{equation}
  C(r,t)=1-\frac{ar}{\sqrt{t}}+\mathcal{O}(t^{-m}),
\end{equation}
where $m>1/2$ and $a=2/\sqrt{\pi}$ for 1D and $a=\kappa$ for 2D.
In order to relate the correlation between sites separated by a distance $nr$, $C(nr,t)$ with $n\geq 1$, and $C(r,t)$, consider the relation
\begin{align}
  [C(r,t)]^n =\left(1-\frac{ar}{\sqrt{t}}+\dots\right)^{\! n}
  =1-n\frac{ar}{\sqrt{t}}+\mathcal{O}(t^{-k}),
\end{align}
with $k>1/2$, that is valid up to ${\cal O}(t^{-1/2})$, the dominant order for $t\gg 1$.
Therefore, up to this order, 
\begin{equation}
  C(nr,t)\approx \left[ C(r,t)\right]^n,
\end{equation}
which, along with the simulations in Fig.~\ref{fig.nnn}, justifies the {\it ansatz} $C_{i,i+2\delta}\approx C_{i,i+\delta}^q$. 
In relation to nearest-neighbors ($r=1$), next-nearest neighbors are located at a distance $nr=2$ in 1D and  $nr=\sqrt{2}$ in 2D. 
However, within our approximation, the stability of the consensus solution is constrained
by the values of $q$ obeying Eq.~(\ref{eqq}). Thus, instead of $\sqrt{2}$, $q=4/3$ is the optimal choice to ensure stability and remaining as close as possible to the correct exponent (see Appendix~\ref{ap.chi-square}).

\section{Conclusions}
\label{sec.conclusions}

In the PVM~\cite{LaDaAr22,LaDaAr24}, each opinion may present two levels of confidence, extreme (zealots) or  regular (normal voters). Zealots are agents that temporarily resist changing their state, regardless of their neighbors' influence.  
The original, more general version of the PVM features non-Markovian transition rules, adding complexity to the analysis of its dynamical evolution.
In the present work we have studied the coarsening dynamics of a simplified version of the PVM, where agents can change their level of confidence at each step based on interactions with their nearest neighbors. Our results indicate that this model retains the key features of the original non-Markovian PVM.  
We derived the governing equations for the one- and two-point correlation functions. Since these equations do not form a closed set, we applied approximate closure schemes in both $d=1$ and $d=2$, whose validity was confirmed through numerical simulations. The analytical solutions obtained from these equations show excellent agreement with the numerical results, further supporting the effectiveness of our approach.

In both the VM and the IM0 the average cluster size $\ell(t)$ increases as the square root of time.
Nonetheless, in $d>1$ their interfaces have different behaviors, with the density of active sites slowly decreasing in time for the VM, $\rho\propto 1/\ln t$.
The PVM, either markovian or not, instead behaves similarly to the IM0, 
with $\rho \propto t^{-1/2}$.
This has been confirmed by the analytical treatment of this paper (numerically, however, small deviations have been spotted, the measured exponent being slightly below 1/2, what was also observed in similar models~\cite{CaEgSa06,DaGa08,VeVa18,MuBiPa20}). The above suggests that the PVM might be in the same universality class of the IM0 (despite, clearly, 
the exponents for $\rho$ and $\ell$
might not be sufficient to discriminate between different universality classes).
If one gets convinced of this, 
the analytic approach for the PVM developed in this paper could also help understanding the kinetics of the IM0 for which, except in $d=1$, analytical approaches are scarce. 

Another possible research direction 
could be the study of the PVM in $3d$. 
Preliminary results indicate that the approximations discussed here, in particular Eq.~(\ref{eq.rhod}), well describe the simulations in this case. 
The application of the PVM to study the ordering kinetics in systems with extended interactions, 
i.e. couplings beyond nearest neighbors, could also be an interesting playground because the ordering kinetics of Ising models with such interactions are not fully understood beyond $d=1$~\cite{PhysRevE.49.R27,christiansen2019phase,christiansen2020aging,PhysRevE.50.1900,corberi2019one}.


\begin{acknowledgments}
Partial funding by the Brazilian Conselho Nacional de Desenvolvimento Científico e Tecnológico (CNPq), Grants 316628/2021-2 (JJA), 443517/2023-1 (JJA) and 402487/2023-0 (WGD and JJA), is acknowledged.

\end{acknowledgments}

\appendix
\section{Symmetry and translational invariance properties}
\label{ap.invariance}

For random initial conditions (null magnetization), the dynamics is invariant under spin flips $S_i \to -S_i$. 
Therefore, correlation functions containing an odd number of $S$ variables should vanish, i.e.
\begin{equation}
C^{j_1, \ldots, j_m}_{i_1, \ldots, i_{2 n+1}} \ = \ 0 \, . 
\label{eq.zerocorr}
\end{equation}
Moreover, the correlations Eq.~(\ref{eq.repcor}) are also invariant under separate permutations of the lower and upper indices
\begin{equation*}
    C_{i_1,\ldots, i_n}^{j_1,\ldots, j_m}  = \  C_{p_i(i_1,\ldots,i_n)}^{p_j(j_1,\ldots,j_m)}  \, . 
\end{equation*}
Finally, translational invariance implies the following useful equalities:
\begin{align}\label{sim1}
& C_{i_1, \ldots, i_n}^{i_j} \ =\  C_{i_1, \ldots, i_n}^{i_k} \, , \quad j,k \in [1,n] \, . \\[2mm]
& \sum _{\delta} C_{i_1, \ldots, i_l+\delta, \ldots, i_n}^{i_k, j_1, \ldots, j_m} \ =\  \sum _{\delta}  C_{i_1, \ldots, {i_k}+\delta, \ldots, i_n}^{i_l, j_1, \ldots, j_m} \, , \\[2mm]
		\label{csim}
& \sum _{\delta} C_{i+\delta,j}  =  \sum _{\delta} C_{i,j+\delta} \, ,\\[2mm]
& \sum_{\delta}C_{j, i +\delta}^i  =  \sum_{\delta}C_{i,j+\delta}^j \, , \label{csim2}
\end{align}


\section{Discrete Laplacian operators}
\label{ap.laplacian}

It can be argued that, in both 1D and 2D, the expressions in Eq.~\eqref{thelaplacians} are indeed discretized Laplacians.
We define $C(r) \equiv C_{i,j}$ and $C^i(r) \equiv C^i_{i,j}$, with $r$ being the distance between $i$ and $j$ (the time dependence is omitted for simplicity). 
Thus, in the 1D case,
\begin{align*}
-C_{i,j}+\frac{1}{2}\sum_{\delta}C_{i+\delta,j} & = -C(r)+\frac{1}{2}\sum_{\delta=\pm 1}C(r+\delta)  \\[2mm]
-C^i_{i,j}+\frac{1}{2}\sum_{\delta}C^i_{i+\delta,j} & = -C^i(r)+\frac{1}{2}\sum_{\delta=\pm 1}C^i(r+\delta) \, .
\end{align*}
For $r \gg 1$, the terms on the r.h.s. can be expanded up to second order, obtaining
\begin{align*}
-C_{i,j}+\frac{1}{2}\sum_{\delta}C_{i+\delta,j} & \approx  \frac{1}{2} \frac{d^2C(r)}{d r^2}   \\[2mm]
-C^i_{i,j}+\frac{1}{2}\sum_{\delta}C^i_{i+\delta,j} & \approx   \frac{1}{2} \frac{d^2 C^i(r)}{d r^2} \, .
\end{align*}
Analogously, in 2D:
\begin{align*}
\sum_{\delta}C_{i+\delta,j}^i  &=   
C_{(i_x+1,i_y),j}^i+C_{(i_x-1,i_y),j}^i \\ 
& +C_{(i_x,i_y+1),j}^i+C_{(i_x,i_y-1),j},
\end{align*}
where each index is a vector in $\mathbb{R}^2$. For simplicity we take $j=(0,0)$ then $r^2 = i_x^2+i_y^2$. Thus
\begin{align*} 
C^i(r)&-\frac{1}{4} \left[C^i\left(\sqrt{(i_x+1)^2+i_y^2}\right)+C^i\left(\sqrt{(i_x-1)^2+i_y^2}\right)\right. \\
&\left. +C^i\left(\sqrt{i_x^2+(i_y+1)^2}\right)+C^i\left(\sqrt{i_x^2+(i_y-1)^2}\right)\right] \\[2mm]
& \approx -\frac{1}{4}\left[ \frac{1}{r} \frac{d C^i(r)}{d r}+\frac{d^2C^i(r)}{d r^2}\right] \ = \ - \frac{1}{4}\Delta C^i(r) \, .
\end{align*}
An analogue expression holds for $C(r)$.

\section{The equality of $\Delta C_{i,j}$ and $\Delta C_{i,j}^i$ in 2D}
\label{ap.equallaplacians}

There are a couple of facts about Fig.~\ref{fig_interf_schem} and the argument around it that must be stressed: i) if we rigidly translate the five marked sites in Fig.~\ref{fig_interf_schem} one lattice spacing to the right (so that $i\to i'$ crosses the interface) $\Delta C_{i,j}$ does not change ($\Delta C_{i,j}^i$ can at most change sign, depending on the state of the variables $\theta $ on the interfacial points $i,i'$). ii) If we move only the site $i$ by one lattice spacing to the right (keeping the top and bottom $j$ sites fixed) we are computing a different Laplacian $\Delta C_{i',j}$, because the distance $|i'-j|$ between $i'$ and the two $j$s along the vertical is slightly increased with respect to $|i-j|$. However, for large distances $|i-j|$ this difference is small and one has
$\Delta C_{i,j}\simeq -\Delta C_{i',j}$.
iii) $\Delta C_{i,j}$ does not change even if we rigidly translate the five marked points vertically sat at one lattice spacing, letting $i\to i''$. Instead, $\Delta C_{i,j}^i$ can change sign if the state of the variable $\theta$ on $i$ and $i''$ is different. This could lead us to the (wrong) conclusion that the contribution to the two Laplacians at fixed $|i-j|$ constructively sum up
in $\Delta C_{i,j}$, when we translate along the surface, and destructively
in $\Delta C_{i,j}^i$. However this is not true because when $\Delta C_{i,j}^i$ changes sign in moving vertically $i$ to $i''$, it adds a contribution $\Delta C_{i'',j}^{i''}$ which is almost equal, and of the same sign, as the contribution $\Delta C_{i',j}$ discussed at point ii). Hence we can conclude that, to a first order approximation, $\Delta C_{i,j}$
and $\Delta C_{i,j}^i$ are equal (for fixed $|i-j|$). 

\section{The closure $C_{i,i+2\delta} \simeq C_{i,i+\delta}^q$ and the value of $q$}
\label{ap.chi-square}

We give some numerical details on the first closure approximation used in Sec.~\ref{sec.stat},  $C_{i,i+2\delta} \simeq C_{i,i+\delta}^q$.
In Fig.~\ref{figxi} we plotted the residual sum of squares,
\begin{equation*}
    \xi = \sum_{i=1}^N[y_i-f(x_i)]^2,
\end{equation*}
where $f(x)=x^q$ and $x_i$ and $y_i$ are, respectively, the values of $C_{i,i+\delta}$ and $C_{i,i+2\delta}$ in the 2D case.
The best fit for this function is obtained using $q=\sqrt{2}$ and not $q=4/3$. 
In fact, the values $q=\sqrt{2}$ for 2D and $q=2$ for 1D suggest that $q$ is the distance between next-nearest neighbors sites. 
However, $q=\sqrt{2}$ in 2D does not lead to the expected $\rho=0$ solution when the system reaches consensus. Among the values of $q$ compatible with such solution, the best fit is obtained using $q=4/3$. 

\begin{figure}[ht!]
    \includegraphics[width=\columnwidth]{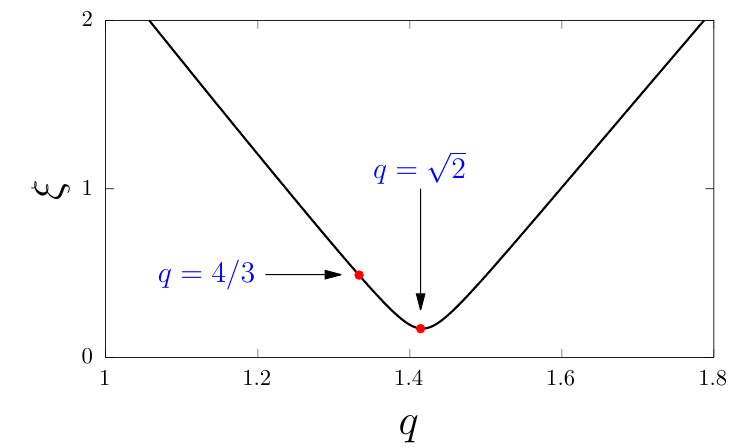}
    \caption{Residual sum of squares for a fit using $f(x)=x^q$ and $x_i$ and $y_i$ as the values of $C_{i,i+\delta}$ and $C_{i,i+2\delta}$, respectively, for the 2D case. The minimum residue  $q=\sqrt{2}$ and the value used in our approximation, $q=4/3$, are highlighted.}
    \label{figxi}
\end{figure}

%

\end{document}